# Long-range micro-pulse aerosol lidar at 1.5 μm with an up-conversion single-photon detector

Haiyun Xia,[1,3] Guoliang Shentu,[2,4] Mingjia Shangguan,[1] Xiuxiu Xia,[2] Xiaodong Jia,[1] Chong Wang,[1] Jun Zhang,[2] Qiang Zhang,[2,4,*] Xiankang Dou,[1,4] and Jianwei Pan[2,4]

[1]*School of Earth and Space Sciences, University of Science and Technology of China, Hefei, 230026, China*
[2]*Shanghai Branch, National Laboratory for Physical Sciences at Microscale and Department of Modern Physics, University of Science and Technology of China, Shanghai, 201315, China*
[3] *Collaborative Innovation Center of Astronautical Science and Technology, HIT, Harbin 150001, China*
[4] *Synergetic Innovation Center of Quantum Information and Quantum Physics, USTC, Hefei 230026, China*
*Corresponding author: qiangzh@ustc.edu.cn*

A micro-pulse lidar at eye-safe wavelength is constructed based on an up-conversion single-photon detector. The ultralow noise detector enables using integration technique to improve the signal-to-noise ratio of the atmospheric backscattering even at daytime. With the pulse energy of 110μJ, the pulse repetition rate of 15 kHz, the optical antenna diameter of 100 mm and integration time of 5 minutes, a horizontal detection range of 7 km is realized. In the demonstration experiment, atmospheric visibility over 24 hours is monitored continuously, with results in accordance with the weather forecasts.

As a minor constituent of the atmosphere, aerosols not only play an important role in the radiation budget, air chemistry and hydrological cycle, but also affect the public health, air visibility and the environment quality. The occurrence, residence time, physical properties and chemical composition verify fast in space and time due to variable aerosol sources and meteorological processes. Lidar is recognized as the only method, which can provide global height-resolved observation of the aerosols [1].

For full-time automatic operation of lidars in the field, the issue of eye safety should be considered. Fortunately, the 1.5 μm laser shows the highest maximum permissible exposure in the wavelength ranging from UV to NIR [2]. What is more, in contrast with UV and Visible systems, 1.5 μm lidars also suffer lower atmospheric attenuation and lower disturbance from Rayleigh scattering and sky radiance, improving the signal-to-noise ratio (SNR).

During clear weather conditions, the pulse energy of the aerosol should be higher than 100 mJ to guarantee the detection range of 5 to 10 km [3]. Obviously, the peak power of such a laser pulse is far beyond the stimulated Brilliouin scattering (SBS) threshold of any erbium doped fiber amplifier (EDFA) [4]. Thus, complex and expensive solid-state techniques including stimulated Raman scattering (SRS) [5], optical parametric oscillator [6], and resonator with Cr[4+]: YAG crystal [7] are developed to generate high power lasers at 1.5 μm for aerosol lidars.

In principle, the SNR of a lidar is dominated by the product of laser power and the telescope area. Although the detectors used in a lidar finally decide the quality of the raw data, their contribution is often neglected. Because the quantum efficiency is simply treated as a constant once the cathode material of a detector is chosen according to a specific working wavelength. For example, InGaAs avalanche photodiode (APD) is used for 1.5μm detection commonly. But it suffers from low efficiency (about 18%), high noise (a few kHz) and high afterpulsing probability due to impurities and defects of the cathode material [8]. Superconductor detector at optical communication band has the advantages of high efficiency (about 25%), ultralow noise (about 10 Hz) and high time resolution (timing jitter of 60 ps) [8]. However, the requirement for liquid helium refrigeration equipment restrains their practical applications in many fields.

Generally, traditional micro-pulse lidar is thought quite unsuitable for long-range aerosol detection with high temporal/spatial resolution [3]. In this work, a compact micro-pulse aerosol lidar incorporating a fiber laser at 1.5 μm, a small optical antenna and an up-conversion detector (UCD) is proposed for long-range aerosol detection during day and night, as shown in Fig. 1.

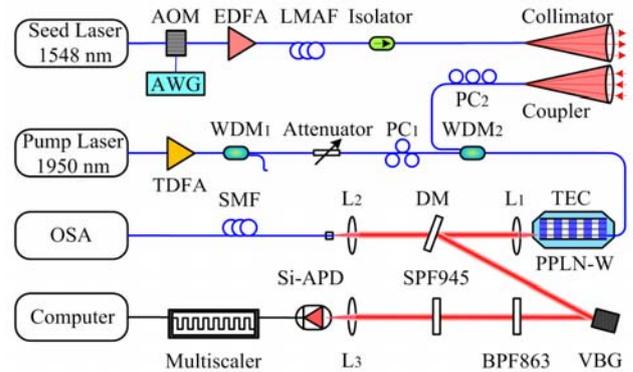

Fig. 1. System layout of the infrared aerosol lidar. AOM, acousto-optic modulator; AWG, arbitrary waveform generator; TDFA, thulium doped fiber amplifier; LMAF, large mode area fiber; SMF, single mode fiber; WDM, wavelength division multiplexer; PC, polarization controller; OSA, optical spectrum analyzer; DM, dichroic mirror; L, lens; SPF, short pass filter; BPS, band pass filter; VBG, volume Bragg grating; TEC, thermoelectric cooler.

The UCD up-converts photons at communication band to visible photons when quasi-phase matching (QPM) is achieved in a periodically poled lithium niobate waveguide (PPLN-W). Then single photons at 1.5 μm can be counted by using a Si: APD with high efficiency and low noise [9].

At present, commercially available fiber lasers benefit from the reliable optic components widely used in the telecommunications industry. Here, the continuous wave from a seed laser (Keyopsys, PEFL-EOLA) is chopped into pulse train after passing through an acousto-optic modulator (Gooch & Housego, T-M080) driven by an arbitrary waveform generator (Agilent, 33250A), which determines the shape of the laser pulse and its repetition rate. The weak laser pulse is fed to a polarization maintaining EDFA (Keyopsys, PEFA-EOLA), which delivers pulse train with energy up to110µJ. A large mode area fiber with numerical aperture of 0.08 is used to enhance the threshold of SBS and avoid self-saturation of amplified spontaneous emission (ASE). Then, the laser is collimated and sent to the atmosphere. It is worth a mention that the maximum permissible ocular exposure to the output laser beam is 2554 mJ [2], which is far beyond the pulse energy of 110 µJ we used here.

The backscattering from the atmosphere is coupled into a single mode fiber by using a pigtail coupler. It should be noted that the micro-pulse transmitter at 1.5 µm with such a small optical antenna are generally used for object ranging with time-correlated single photon counting technique [10] or atmosphere wind detection using heterodyne technique [11]. Direct detection of the weak aerosol backscattering signal is a great challenge for micro-pulse lidar. In this work, an up-conversion single-photon detector with high detection efficiency and ultralow dark count noise is used. The continuous wave from the pump laser at 1950 nm is followed by a thulium-doped fiber amplifier, both manufactured by AdValue Photonics (Tucson, AZ). The residual ASE noise is suppressed by using a 1.55/1.95-µm wavelength division multiplexer (WDM1). The atmosphere backscattering and the pump laser are combined and transferred into a PPLN-W via the second WDM2. Optimized QPM condition is guaranteed by tuning the temperature of the PPLN-W with a thermoelectric cooler. Then backscattering photons at 1548 nm are converted into sum-frequency photons at 863 nm, which is collimated through an AR-coated objective lens and picked out by using a dichroic mirror. A volume Bragg grating (VBG)in conjunction with an 863 nm band pass filter (BPF) and a 945 nm short pass filter (SPF) is used to block the second and higher order harmonics of the pump and the spontaneous Raman scattering. Finally the 863 nm signal is focused onto a Si-APD. The TTL signal corresponding to the photons received on the Si-APD (PicoQuant, τ SPAD) is fed to a multiscaler (FAST ComTec, MCS6A) and then processed in the computer. The optical layout of the up-conversion detector is shown in Fig. 2. For reader's convenience, key parameters of the micro-pulse aerosol lidar are listed in table 1.

In addition to the advantages inherited from the detection laser at 1.5 µm, the micro-pulse lidar based on the up-conversion detector has some other outstanding features for daytime operation. Firstly, the waveguide only supports TM-polarized light, so both the polarization states of the pump laser and the atmosphere backscattering should be controlled. This will suppress the sky radiance by a factor of 2. Secondly, the usage of VBG, BPF and SPF in sequence results an ideal optical filter with background suppression>55dB, bandwidth of 0.05nm and transmission of 88 %. Thirdly, thanks to the quality of the Si: APD, a detection efficiency of 15 % with the dark count noise of 40 Hz is realized at 1.5 µm, when the power of the 1950 nm pump laser is tuned to 25 mW [9], which allows integration of ultralow backscattering signal over a preset number of laser pulses emitted by the micro-pulse lidar.

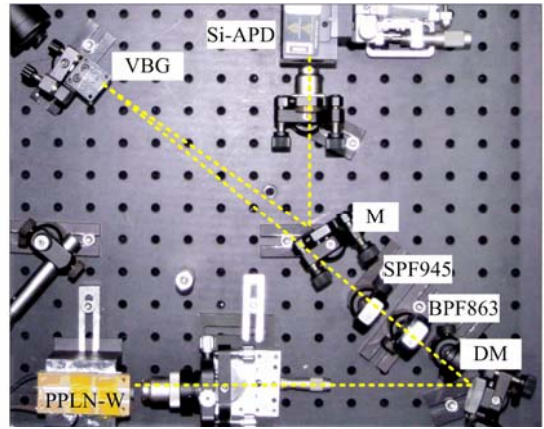

Fig. 2.Picture of the up-conversion single-photondetector.

Table.1 Key parameters of the infrared aerosol Lidar

| Parameter | Value |
|---|---|
| Pulsed Laser | |
|     Wavelength (nm) | 1548.1 |
|     Pulse duration(ns) | 300 |
|     Pulse energy (µJ) | 110 |
|     Pulse repetition rate (kHz) | 15 |
| Pump Laser | |
|     Wavelength (nm) | 1950 |
|     Power (mW) | 800 |
| Collimator | |
|     Aperture (mm) | 100 |
|     Focal length (mm) | 500 |
|     LMAF Mode-field diameter(µm) | 20 |
| Coupler | |
|     Aperture (mm) | 60 |
|     Focal length (mm) | 150 |
|     SMF Mode-field diameter(µm) | 10 |
| PPLN waveguide | |
|     QPM period (µm) | 19.6 |
|     Length (mm) | 52 |
|     Insert loss (dB) | 1.5 |
| Volume Bragg grating | |
|     Reflection efficiency (%) | 95 |
|     Bandwidth (nm) | 0.05 |
| Silicon APD | |
|     Detection efficiency at 863 nm (%) | 45 |
|     Dark count (Hz) | 25 |

As a demonstration of the new micro-pulse lidar at 1.5 µm based on the up-conversion single-photon detector, the prototype system is operated horizontally for continuous measurement of atmospheric visibility, which is not only significant to the public health but also important for all traffic operations and free-space optical communication. Range-resolved recording of the backscattered signal allows the measurement of turbidity and, thus, of the local visibility.

As an example, experiment data over 24 hours is shown in Fig. 3. The temporal resolution and spatial resolution are set to 300 s and 45 m, respectively. One can see from Fig. 3(a) that, even with only 110 μJ pulse energy and receiver diameter of 60 mm, the raw backscattering signal can extend to 7 km horizontally. During the experiment, from 20:00, Oct. 26, 2014 to 2:00 next morning, smoke from the chimney of a power plant located 6.5 km away is detected four times. One example of the smoke trace is shown in Fig. 3(b).

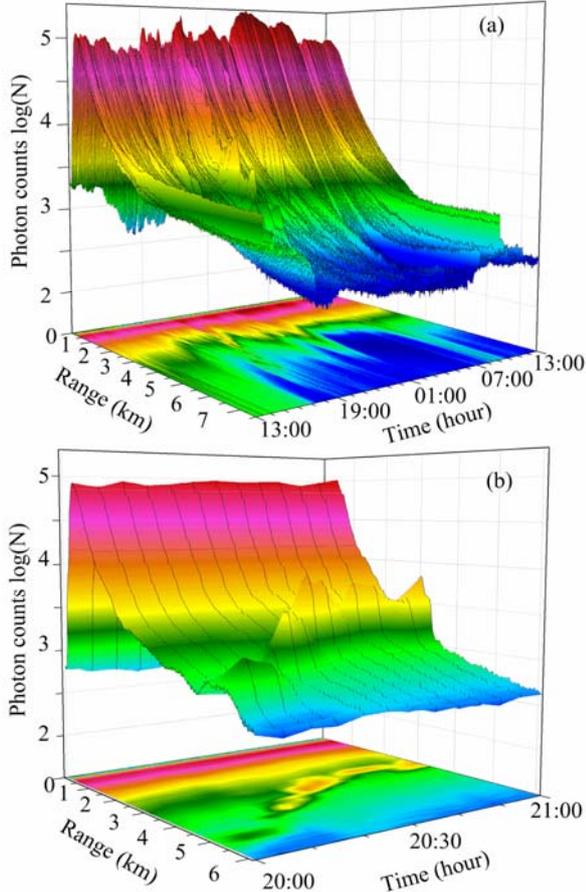

Fig. 3. Experiment raw data from lidar. (a)Backscattering signal of 24-hour continuous horizontal detection start at 13:00 on Oct. 26, 2014. (b) An example of smoke detected in the sight of the lidar over one hour.

Note that Rayleigh scattering is proportional to $\lambda^{-4}$, which is ignored in the atmospheric backscattering by taking advantage of 1.5μm working wavelength. Thus, the photon number corresponding to the aerosol backscattering is described as

$$N(R) = E\eta_o \frac{\eta_q}{h\nu} \frac{A_0}{R^2} \xi(R)\beta(R)\frac{c\tau_d}{2}\exp\left[-2\int_0^R \sigma(R)dR\right] \quad (1)$$

where, $E$ is the energy of the laser pulse, R is the range. $\eta_0$ accounts for the optical efficiency of the transmitted signal, $\eta_q$ is the quantum efficiency, $h$ is the Planck constant, $A_0$ is the area of the telescope, $\xi(R)$ is the geometrical overlap factor, $\beta(R)$ is the Mie volume backscattering coefficients, $\tau_d$ is the detector's response time $\sigma(R)$ is the atmospheric extinction coefficient. The geometrical overlap is calibrated in advance. By hypothesizing a constant ratio between $\sigma(R)$ and $\beta(R)$, an inversion algorithm proposed by Klett and Fernald is applied to retrieved $\sigma(R)$ [1].

The relation between atmospheric visibility and the atmospheric attenuation coefficient is dependent on the physical characteristics of the hydrometeors as well as on the wavelength. According to Isaac I. Kim's model [12], the relation is presented as

$$\sigma = \frac{3.91}{V}\left(\frac{\lambda}{550nm}\right)^{-q} \quad (2)$$

V is the atmospheric visibility in unit of km, $\lambda$ is the detection wavelength in unit of nm, $q$ is the size distribution of the scattering particles, given by

$$q = \begin{cases} 1.6 & (V > 50\ km) \\ 1.3 & (6km < V < 50\ km) \\ 0.16V + 0.34 & (1km < V < 6\ km) \\ V - 0.5 & (0.5km < V < 1km) \\ 0 & (V < 0.5\ km) \end{cases} \quad (3)$$

Finally, atmospheric visibility is retrieved according to Eq. (2), as shown in Fig. 4. The atmospheric humidity and temperature are recorded for reference.

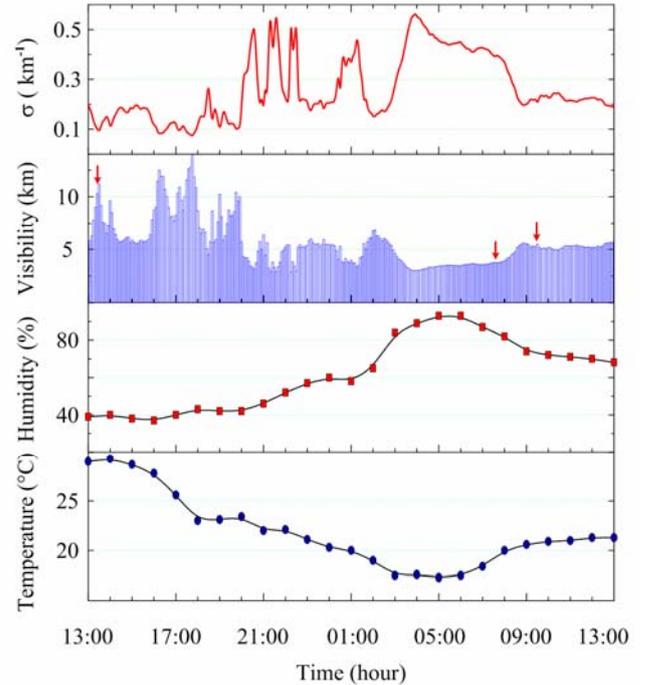

Fig. 4. Experiment results, from top to bottom, are extinction coefficient, visibility, Humidity and atmospheric temperature near ground.

The first peak of atmospheric attenuation coefficient occurs at the rush hour (18:00), which is due to the vehicle exhaust during the traffic congestion. The four bursts followed are due to the smoke in the sight of the optical antenna, as pointed out in Fig .3. Between 2:00 and 8:00, the attenuation coefficient rises fast as the emergence of haze. After the sunrise, the attenuation coefficient declines as the drop/rise of the humidity/temperature. As shown in Fig. 5, the visibility values provided by the municipal meteorological bureau show good correspondence with the lidar detection.

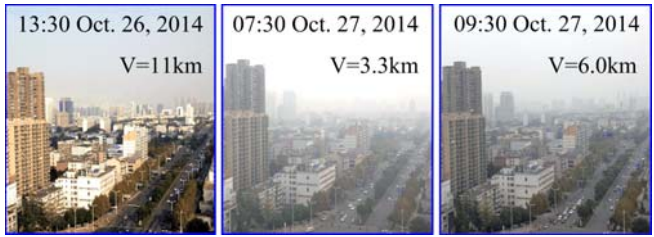

Fig. 5.Photographs of the scene in situ under different weather conditions during the 24-hour lidar detection

After the visibility experiment, the UCD is compared to a commercial InGaAs APD (AUREA, APD-A), which operates in Geiger-mode at 20 MHz. Its quantum efficiency is adjustable by tuning the reverse voltage to the APD. For one minute integral time, the backscattering signals are plotted in Fig. 6. Obviously, the signal-to-noise ratio can be enhanced two orders by using the up-conversion detector instead of the InGaAs APD.

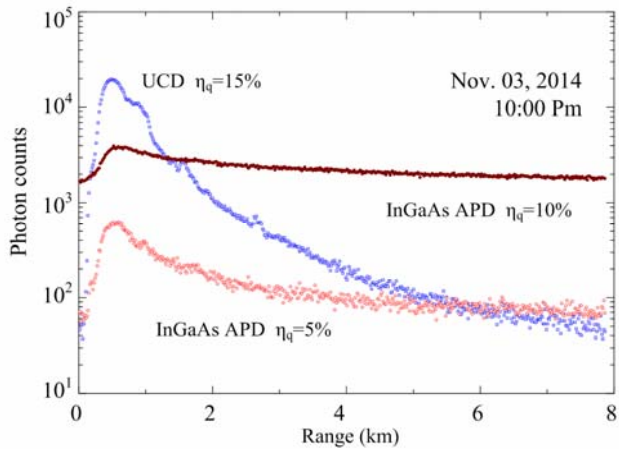

Fig.6.Performance comparison between the InGaAs APD and the up-conversion detector.

In order to test the stability of our setup, the residual signal laser is monitored using an optical spectrum analyzer (YOKOGAWA, AQ6370C) to test the conversion efficiency. Then a tunable C-band Laser is used to achieve the wavelength tuning curve of the PPLN-W when the QPM temperature is 37.8 C and the pump wavelength is 1950nm. As shown in Fig. 7(a), the conversion efficiency is beyond 90 % while the wavelength drift of the 1548.1 nm laser is less than 0.2 nm. Then, a temperature tuning curve is measured by scanning the temperature of waveguide, as shown in Fig. 7(b). The conversion efficiency can be kept above 90% while the temperature is controlled within $38.3 \pm 0.8$ ℃.

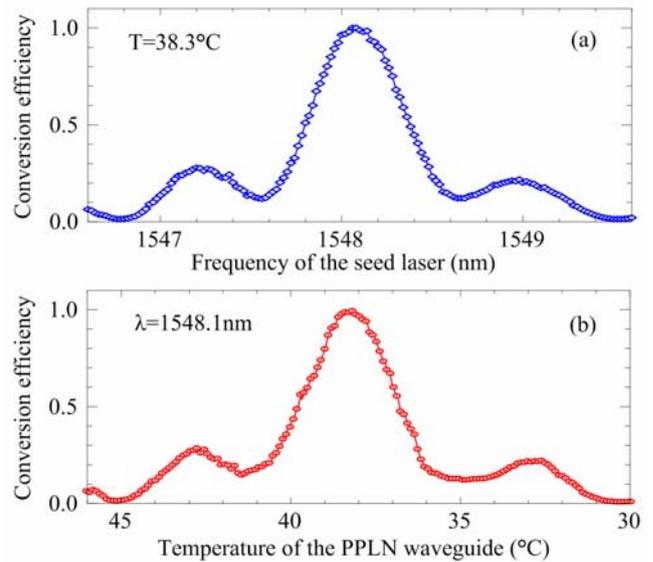

Fig. 7. (a) Conversion efficiency versus the frequency of the 1.5 μm seed laser. (b) Conversion efficiency versus the temperature of the PPLN-W.

In conclusion, for the first time, a long-range micro-pulse aerosol lidar incorporating up-conversion detector was demonstrated for continuous aerosol detection during day and night. In future work, a polarization maintaining and all-fiber up-conversion detector will be developed to improve the polarization stability and compactness of the lidar.

This work has been supported by the NFRP (2011CB921300, 2013CB336800), the NNSF of China (41174131, 41274151), and the CAS.